\documentclass[12pt,preprint]{aastex}
\usepackage{epsfig}

\def\gtrsim{\mathrel{\hbox{\rlap{\hbox{\lower4pt\hbox{$\sim$}}}\hbox{$>$}}}}
\def\lesssim{\mathrel{\hbox{\rlap{\hbox{\lower4pt\hbox{$\sim$}}}\hbox{$<$}}}}

\def\vkm{km s$^{-1}$}

\def\degree{$^\circ$}

\def\arcsa#1#2{$#1^{\prime\prime}_{^\textrm{.}}#2$}

\def\solarmass{$M_\odot$}

\def\mJyb{mJy beam$^{-1}$}

\def\cmc{cm$^{-3}$}

\def\mH2{m_{\textrm{\scriptsize H}_2}}
\def\Ro{R_\textrm{\scriptsize 0}}
\def\ro{r_\textrm{\scriptsize 0}}
\def\rmax{r_\textrm{\scriptsize max}}
\def\rin{r_\textrm{\scriptsize in}}
\def\rout{r_\textrm{\scriptsize out}}
\def\To{T_\textrm{\scriptsize 0}}
\def\no{n_\textrm{\scriptsize 0}}

\def\xSO{x_\textrm{\scriptsize SO}}
\def\texin{\textrm{\scriptsize in}}
\def\texout{\textrm{\scriptsize out}}

\def\H2{H$_2$}
\def\N2HP{N$_2$H$^+$}
\def\HCOP{HCO$^+$}

\def\NH3{NH$_3$}

\def\HCOP{HCO$^+$}

\def\putfiga#1#2#3{\epsfig{scale=#1,angle=#2,figure=#3}}
\def\putfig#1#2#3{}
\def\blank#1{}

\newcounter{mfigure}[section]

\begin{document}

\title{First Detection of Interaction between a Magnetic Disk Wind and an
Episodic Jet in a Protostellar System}

\author{Chin-Fei Lee$^{1,2}$, Benoit Tabone$^{3,4}$, Sylvie Cabrit$^{4}$,
Claudio Codella$^{5,6}$, Linda Podio$^{5}$, Ferreira, J.$^{6}$,
Jacquemin-Ide, J.$^{6}$}

\altaffiltext{1} {Academia Sinica Institute of Astronomy and Astrophysics, No.  1, Sec. 
  4, Roosevelt Road, Taipei 10617, Taiwan}
\altaffiltext{2} { Graduate Institute of Astronomy and Astrophysics, National Taiwan 
   University, No.  1, Sec.  4, Roosevelt Road, Taipei 10617, Taiwan}
\altaffiltext{3} {Leiden Observatory, Leiden University, PO Box 9513, 2300 RA Leiden, The Netherlands }
\altaffiltext{4} {Observatoire de Paris, PSL University, Sorbonne Universit\'e, CNRS, LERMA, 61 Av. de l'Observatoire, 75014 Paris, France}
\altaffiltext{5} {INAF, Osservatorio Astrofisico di Arcetri, Largo E. Fermi 5, 50125 Firenze, Italy}
\altaffiltext{6} {Univ. Grenoble Alpes, CNRS, Institut de Plan\'etologie et d'Astrophysique de Grenoble (IPAG), 38000 Grenoble, France}


\begin{abstract}

Rotating outflows from protostellar disks might trace extended
magneto-hydrodynamic (MHD) disk winds (DWs), providing a solution to the
angular momentum problem in disk accretion for star formation.  In the jet
system HH 212, a rotating outflow was detected in SO around an episodic jet
detected in SiO.  Here we spatially resolve this SO outflow into three
components: a collimated jet aligned with the SiO jet, the wide-angle disk
outflow, and an evacuated cavity in between created by a large jet-driven
bowshock.  Although it was theoretically predicted before, it is the first
time that such a jet-DW interaction is directly observed and resolved, and
it is crucial for the proper interpretation and modeling of non-resolved DW
candidates.  The resolved kinematics and brightness distribution both
support the wide-angle outflow to be an extended MHD DW dominating the local
angular momentum extraction out to 40 au, but with an inner launching radius
truncated to $\gtrsim 4$ au.  Inside 4 au, where the DW may not exist, the
magneto-rotational instability (MRI) might be transporting angular momentum
outwards.  The jet-DW interaction in HH 212, potentially present in other
similar systems, opens an entirely new avenue to probe the large-scale
magnetic field in protostellar disks.

\end{abstract}


\section{Introduction}

Rotating outflows from protostellar disks are a newly discovered component
in star formation
\citep{Launhardt2009,Greenhill2013,Zapata2015,Bjerkeli2016,Hirota2017,Tabone2017,Lee2018Dwind,Lee2018HH211,Zhang2018,Louvet2018,deValon2020}. 
They might trace extended magneto-hydrodynamic (MHD) disk winds (DWs),
providing a solution to the angular momentum problem in disk accretion.  The
HH 212 protostellar system \citep{Zinnecker1992} is a young system located
in Orion at $\sim$ 400 pc, harboring a nearly edge-on rotating disk optimal
for detecting such an extended disk wind.  The central protostar has a mass
of 0.25$\pm$0.05 \solarmass{}, deeply embedded in an infalling-rotating
envelope \citep{Lee2017COM}.  Previous Atacama Large
Millimeter/submillimeter Array (ALMA) observations have spatially resolved
the disk \citep{Lee2017Disk}, which is rotating within a centrifugal barrier
at $\sim$ 44 au \citep{Lee2017COM}.
 A spinning jet was also detected in SiO carrying angular momentum away from
the innermost disk at a radius of $\sim$ 0.05$-$0.20 au
\citep{Lee2017Jet,Tabone2017}, allowing disk material to fall onto the
protostar.  Dust polarization was also detected towards the disk, suggesting
a presence of a poloidal magnetic field that could launch a disk wind around
the jet \citep{Lee2018Bdisk}.

A slow wide-angle outflow was indeed detected in SO and SO$_2$ at 60 au
resolution, rotating in the same sense as the disk, and consistent with an
extended MHD disk wind launched from $\simeq 0.1$ to 40 au that extracts
enough angular momentum to drive disk accretion \citep{Tabone2017}.  Later
observations at $\sim$ 4 times higher resolution resolved SO emission into a
collimated jet and a wide-angle rotating shell \citep{Lee2018Dwind}.  Now
with new ALMA observations about two times deeper and a spatial resolution
of $\sim$ 13 au (\arcsa{0}{033}), we retrieve additional SO emission
structures reconciling the above seemingly contradictory results, providing
a confirmation for an extended disk wind as well as the first evidence of
jet-disk wind interaction first predicted by \citet{Tabone2018}.  This
interaction provides unique first clues to the unknown magnetic field
strength and distribution in young protostellar disks.


\section{Observations} 

HH 212 was observed with ALMA in Band 7 centered at a frequency of $\sim$
341.5 GHz on 2017 November 27 in Cycle 5 (Project ID: 2017.1.00044.S). 
Since the observations and calibrations have been reported in
\citet{Lee2019COM}, here we only report important information related to the
SO line at 346.528481 GHz and the SiO line at 347.330631 GHz.  Three
scheduling blocks were executed with an on-source time of 98 minutes.  The
projected baselines were $\sim$60$-$8500 m.  We set up the correlator to
have three continuum windows and one spectral window.  The SiO and SO lines
were both included in the spectral window, which has  a velocity resolution of $\sim$ 1.69
\vkm{} per channel.



The $uv$ data was calibrated manually by the ALMA QA2 team using the Common
Astronomy Software Applications (CASA) package version 5.1.1.  No
self-calibration was performed due to insufficient signal to noise ratio of
the continuum data in the long baselines.  A robust factor of 0.5 was used
for the visibility weighting to generate SO and SiO channel maps with a
synthesized beam of \arcsa{0}{036}$\times$\arcsa{0}{030} at a position angle
of $\sim$ $-$78\degree{} and a noise level of $\sim$ 0.75 \mJyb{} (7.0 K).  We
also included the SO visibility data obtained in Cycle 3
\citep{Lee2018Dwind} and reduced the noise level slightly down to 0.67
\mJyb{} (6.2 K) in the SO channel maps.  The velocities in the
channel maps are LSR velocities.

\section{Results}

\begin{figure}
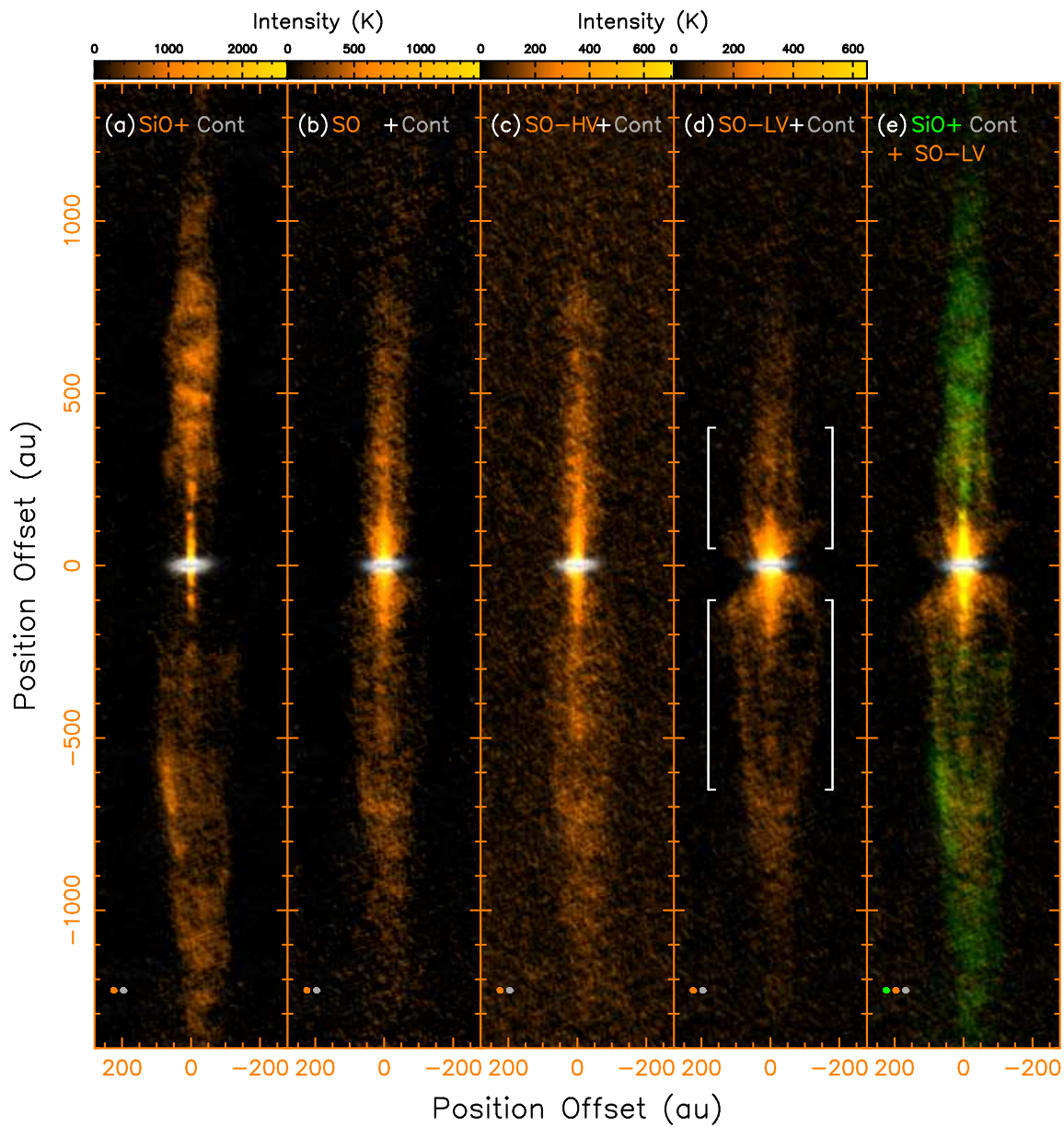
 \centering 
\putfiga{1.1}{270}{f1.eps} 
\figcaption[]
{SiO and SO intensity maps toward the HH 212 system within 1400 au of the protostar,
together with the 350 GHz continuum map of the disk 
\cite[gray image adopted from][]{Lee2019COM}.
The maps are all rotated by 22.5\degree{} clockwise to
align the jet axis in the north-south direction.
SO-HV indicates the SO map at high velocity more than $\pm3$ \vkm{} away from the systemic velocity.
SO-LV indicates the SO map at low velocity within $\pm3$ \vkm{} of the systemic velocity.
Color codes are the same as the labels. White brackets mark the shells.
\label{fig:SiO_SO}}
\end{figure}

Figure \ref{fig:SiO_SO} shows the SO map in comparison to the SiO map of the
jet and the continuum map of the dusty disk \cite[adopted
from][]{Lee2019COM} within 1400 au of the central protostar at 13 au
resolution.  SiO shows an episodic jet launched from the innermost disk,
appearing first as a highly collimated chain of knots in inner 200 au and
then a chain of broader bow shocks downstream at larger distances.  SO also
shows a jet aligned with the SiO jet.

We can unveil the wide-angle outflow by separating the SO emission into two
velocity components.  At high velocity (more than $\pm3$ \vkm{} away from
the systemic velocity of $\sim$ 1.7 \vkm{}, Figure \ref{fig:SiO_SO}c), SO
traces a collimated jet aligned with the SiO jet, but wider possibly because
the SO line has a lower critical density than SiO and thus can trace less
dense material.  The critical densities (in H$_2$) are $7.2\times10^6$
\cmc{} for SO and $1.2\times10^7$ \cmc{} for SiO \citep{Schoier2005}.  At
low velocity (within $\pm$ 3 \vkm{} of the systemic velocity, Figure
\ref{fig:SiO_SO}d), thin outflow shells (marked with white brackets) are
detected in SO surrounding the jet.  Only their bases were detected before
\citep{Lee2018Dwind}.  They are now detected further away and seen to
smoothly connect to the SiO/SO bow shocks downstream at larger distances
($\sim$ 600 au) (Figure \ref{fig:SiO_SO}e).  Faint extended SO emission is
also detected surrounding the base of the shells, within $z \lesssim$ 150 au
from the disk.  This emission shows up better in an intensity-weighted
velocity map (Figure \ref{fig:SOmom1}), forming a wide-angle rotating
outflow together with the base of the shells, appearing as a thick X-shape
fanning out from the disk, rotating around the jet.  Away from the base, the
shells are mainly blueshifted in the north and redshifted in the south,
similar to the velocity sense of the bow shocks at larger distance and thus
driven by them.  The inner part of the wide-angle outflow coincides with the
base of the shells and is thus perturbed by the bow shocks.  The wide-angle
outflow has an outer boundary outlined by the inner infalling-rotating
envelope traced by the high-velocity emission of HCO$^+$ (Figure
\ref{fig:SOmom1}b), confirming that it originates from the disk.  Its outer
part is unperturbed by the bow shocks, providing the best opportunity to
check the previously proposed MHD disk wind interpretation
\citep{Tabone2017,Tabone2020}.

\section{MHD Disk Wind Model}

\def\vkep{v_\textrm{\scriptsize kep}}
\def\vkepo{v_{k,0}}
\def\nmh{n_{\scriptsize H_2}}
\def\vex{v_\textrm{\scriptsize ex}}
\def\ts{t_\textrm{\scriptsize s}}

Various MHD models are being developed to launch disk winds and carry away
part or all of the angular momentum from accretion disks
\citep{Turner2014,Bai2017,Zhu2018,Riols2020}.  The first and most simple 2D
version of these models is a steady-state, axisymmetric, self-similar wind
launched from a geometrically thin Keplerian disk
\citep{Blandford1982,Ferreira1997}.  These models are well suited for
comparison with observations because they allow for parameter studies.
As discussed in \citet{Tabone2020}, the
observable structure and kinematics of the wind in these models are mainly
determined by three parameters: (1) protostellar mass $M_\ast$ defining the
Keplerian rotation $\vkepo=\sqrt{G M_\ast/r_0}$ at a radius $r_0$ in the
disk, (2) magnetic level arm parameter $\lambda \simeq (r_A/r_0)^2$, where
$r_A$ is the Alfv$\acute{\textrm{e}}$n radius along the streamline launched
from a footpoint at $r_0$, determining the poloidal acceleration and the
extracted specific angular momentum.  In particular, the terminal wind
velocity and the final specific angular momentum 
achieved along each streamline are $v_w \sim \vkepo
\sqrt{2\lambda-3}$ and $l \sim \lambda \,l_0$, respectively,
where $l_0= r_0 \,\vkepo$ is the value
at the footpoint at $r_0$.  And (3) widening factor $W \equiv \rmax/r_0$,
where $\rmax$ is the maximum radius reached by the streamline at large
distance, controlling the flow transverse size.


\def\no{n_\mathrm{o}}
\def\na{n_\mathrm{t}}
\def\Ro{R_\mathrm{o}}
\def\Ra{R_\mathrm{t}}
\def\To{T_\mathrm{o}}
\def\Ta{T_\mathrm{t}}
\def\ho{h_\mathrm{o}}
\def\ha{h_\mathrm{t}}

\def\vk{v_\mathrm{ko}}
\def\cs{c_\mathrm{s}}
\def\vko{v_\mathrm{ko}}
\def\cso{c_\mathrm{so}}
\def\vp{v_\phi}
\def\vkep{v_\textrm{\scriptsize kep}}
\def\nmh{n_{\scriptsize H_2}}

\begin{figure}
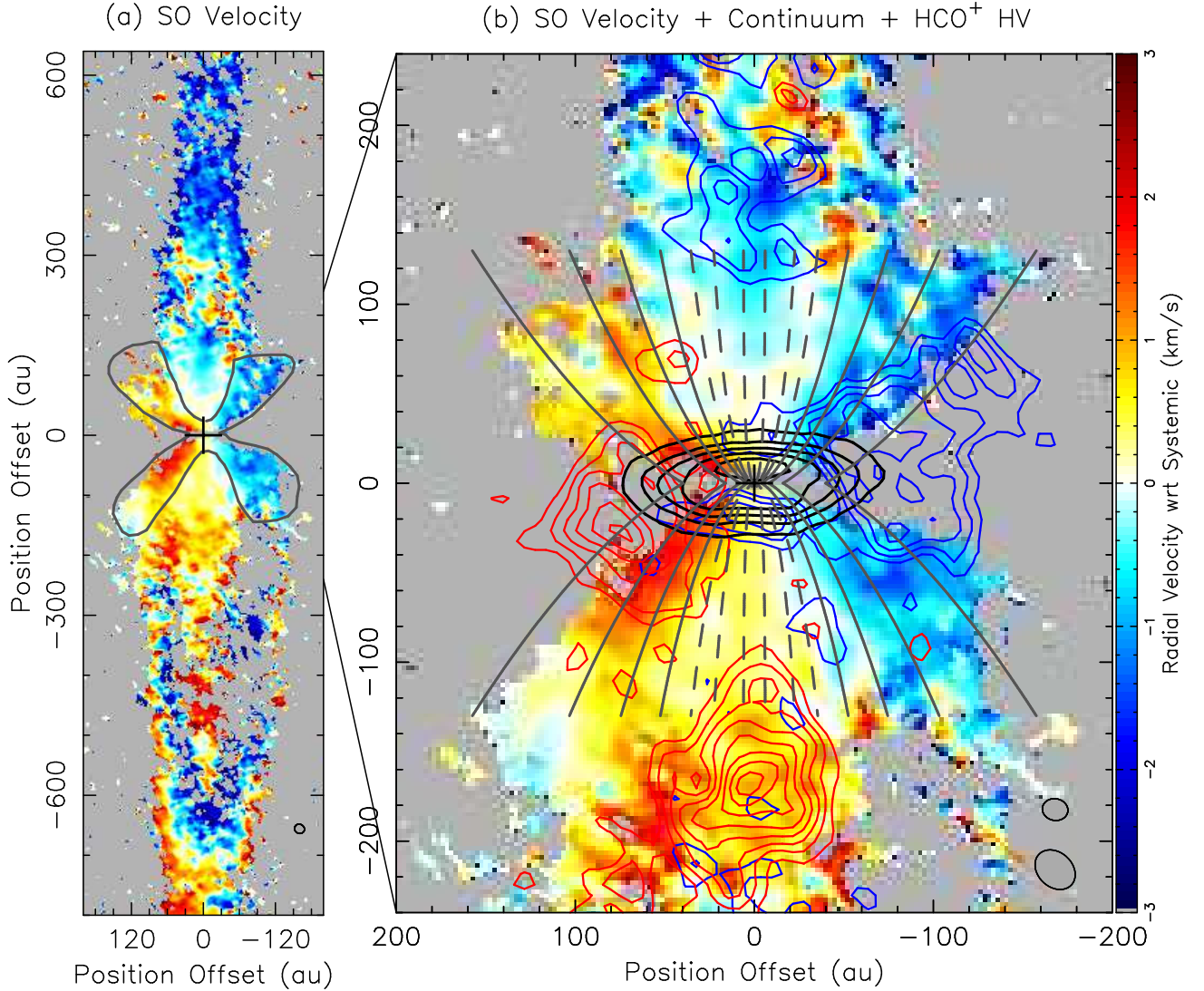
 \centering 
\putfiga{0.8}{270}{f2.eps} 
\figcaption[]
{Intensity-weighted velocity maps of SO in the inner region at low velocity. 
The gray X-shaped curve in (a) outlines the wide-angle rotating outflow
detected in SO.  Panel (b) zooms into the central region.  Black contours
show the same disk map as in Figure \ref{fig:SiO_SO}.
Red and blue contours show the high-velocity (HV) \HCOP{} emission adopted
from \citet{Lee2017COM}, outlining the boundary of the innermost envelope. 
Gray curves plot the streamlines of the disk wind in Model L5W30, with
footpoints at $r_0=$ 4, 8, 16, and 40 au.  Dashed curves show
the streamlines with footpoints at $r_0=$ 0.2, 1, 2 au.
\label{fig:SOmom1}}
\end{figure}

\begin{figure}
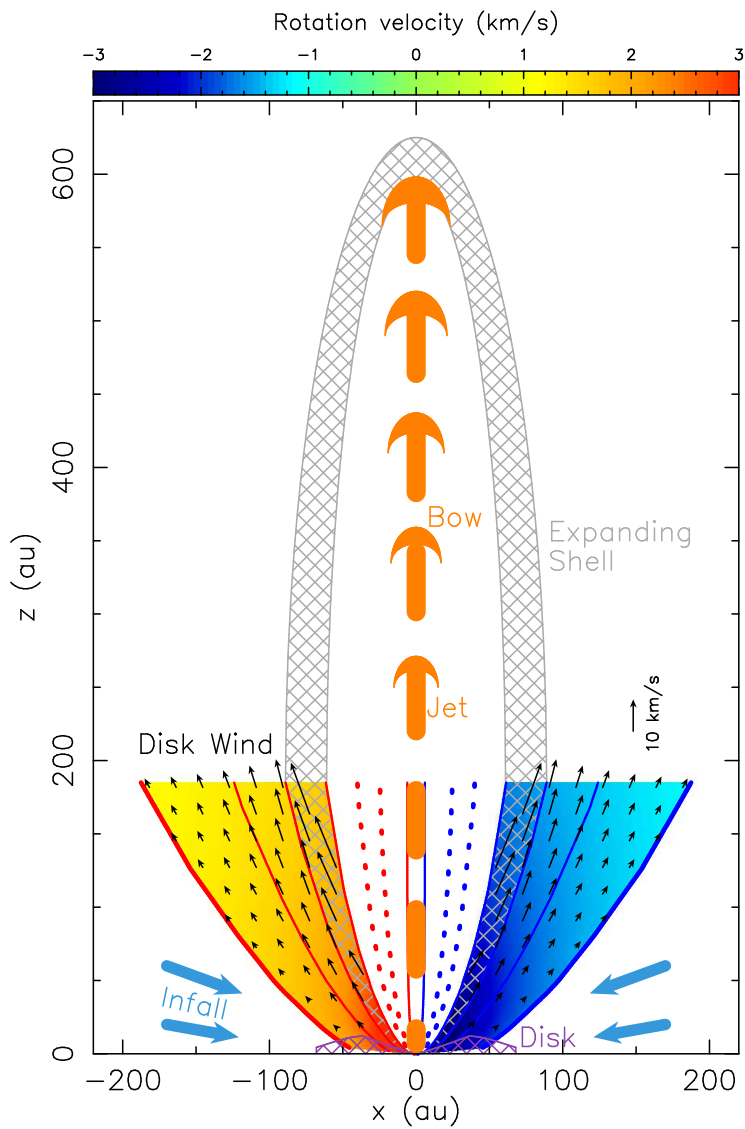
 \centering 
\putfiga{0.8}{270}{f3.eps} 
\figcaption[]
{A schematic diagram showing the extended disk wind and jet in our model,
and the interaction between them.  The wind has a launching radius of 4 to
40 au, while the jet has a launching radius of 0.05 to 0.20 au.  The shell
(gray cross-hatched region) extends from the inner part of the wind to the
jet-driven bow shock.  The rotation velocity (color image) and outflow
velocity (vectors) in the wind are derived from Model L5W30, with the outflow
velocity in the inner
part replaced with a radially expanding velocity.
\label{fig:cartoon}}
\end{figure}

Assuming $M_\ast \sim 0.2$\solarmass{}, \citet{Tabone2017,Tabone2020} found
their disk wind Model L5W30 (with $\lambda \sim 5.5$ and $W \sim 30$) to
broadly reproduce the transverse spatio-kinematic structure of the SO
rotating outflow, with $r_0$ = 0.1 to 40 au.  The same model is thus adopted
here to compare with the SO wide-angle outflow resolved at higher resolution
and sensitivity.  As shown in Figure \ref{fig:SOmom1}b, the wide-angle
outflow shows an opening structure in good agreement with the predicted
model streamlines but with a launching radius truncated to $\gtrsim 4$ au. 
We thus assume an extended disk wind with a launching radius of $r_0 \sim
4-40$ au for the wide-angle outflow, as shown in Figure \ref{fig:cartoon}. 
The wind is assumed to be symmetric with respect to the disk midplane and
extend out to 185 au to the north and south.  The inner part of the wind
bounded by the model streamlines launched from 4 to 8 au roughly coincides
with the base of the shells (Figure \ref{fig:SOmom1}b) and is thus assumed
to become the shell perturbed by the bow shocks, with its outflow velocity
replaced by a radially expanding velocity  $v_r= r/\ts $, where
$\ts$ is the dynamical age of the shell \citep{Lee2018Dwind}.  We assume a
temperature of 100 K in the outer unperturbed part and 200 K in the inner
perturbed part (shell), based on the temperatures derived before for the
disk atmosphere and the shells \citep{Lee2018Dwind}.  The SO jet is assumed
to have a launching radius of 0.10 to 0.20 au in the dust-free zone
\citep{Tabone2020}.  It has a temperature of 300 K, which is a mean value
adopted before to derive the SO abundance \citep{Podio2015}.  Since the
observed jet is unresolved, this jet component is only for illustrative
purposes.

In this self-similar model, the disk has an accretion rate varying with
radius as \begin{equation} \dot{M}_{acc} (r_0) =
\dot{M}_{in}\big(\frac{r_0}{r_{in}}\big)^\xi \end{equation} with
$\dot{M}_{in}$ being the accretion rate at the inner radius $r_{in}$ and
$\xi$ being the ejection efficiency \citep{Ferreira1995}.  Thus, the
mass-loss rate in the wind between the inner radius $\rin$ and outer radius
$\rout$ will be \begin{equation} \dot{M}_{DW}=\dot{M}_\texout-\dot{M}_\texin
= \dot{M}_{in}[(\frac{\rout}{\rin})^\xi-1] \label{eq:Mdw}\end{equation}
where $\dot{M}_\texout$ is the accretion rates at $\rout$.  For the disk
wind to remove all of the accretion angular momentum, we have
\citep{Tabone2020} \begin{equation} \xi\sim \frac{1}{2(\lambda-1)}. 
\label{eq:xi-lambda} \end{equation} With $\lambda \sim 5.5$, $\rout \sim 40$
au, $\rin \sim 4$ au, and $\dot{M}_\texin \sim 3\times10^{-6}$ \solarmass{}
yr$^{-1}$ \citep{Lee2020Jet}, we have $\dot{M}_{DW} \sim 0.9\times10^{-6}$
\solarmass{} yr$^{-1}$.  Along each streamline, the corresponding wind
density has been given in \citet{Tabone2020}.









This model can be compared quantitatively to the observed wide-angle outflow
in terms of kinematics.  A radiative transfer code \citep{Lee2014} adding
the SO line is used to generate the position-velocity (PV) diagrams of the
SO emission from the model, assuming LTE.  The SO abundance (wrt molecular
hydrogen) $\xSO$ is a free parameter to be derived by matching the observed
intensity.  Within 100 au of the protostar, since the jet has a proper
motion of $\sim$ 64 \vkm{} \citep{Claussen1998} and a mean radial velocity
of $\sim$ $-$3.7 \vkm{} in the northern component and $\sim$ 2.5 \vkm{} in
the southern component \citep{Lee2017Jet}, the inclination angles of the
wide-angle outflow and jet are assumed to be $\sim
-3$\degree{} in the northern component and $\sim 2$\degree{} in the southern
component.



\begin{figure}
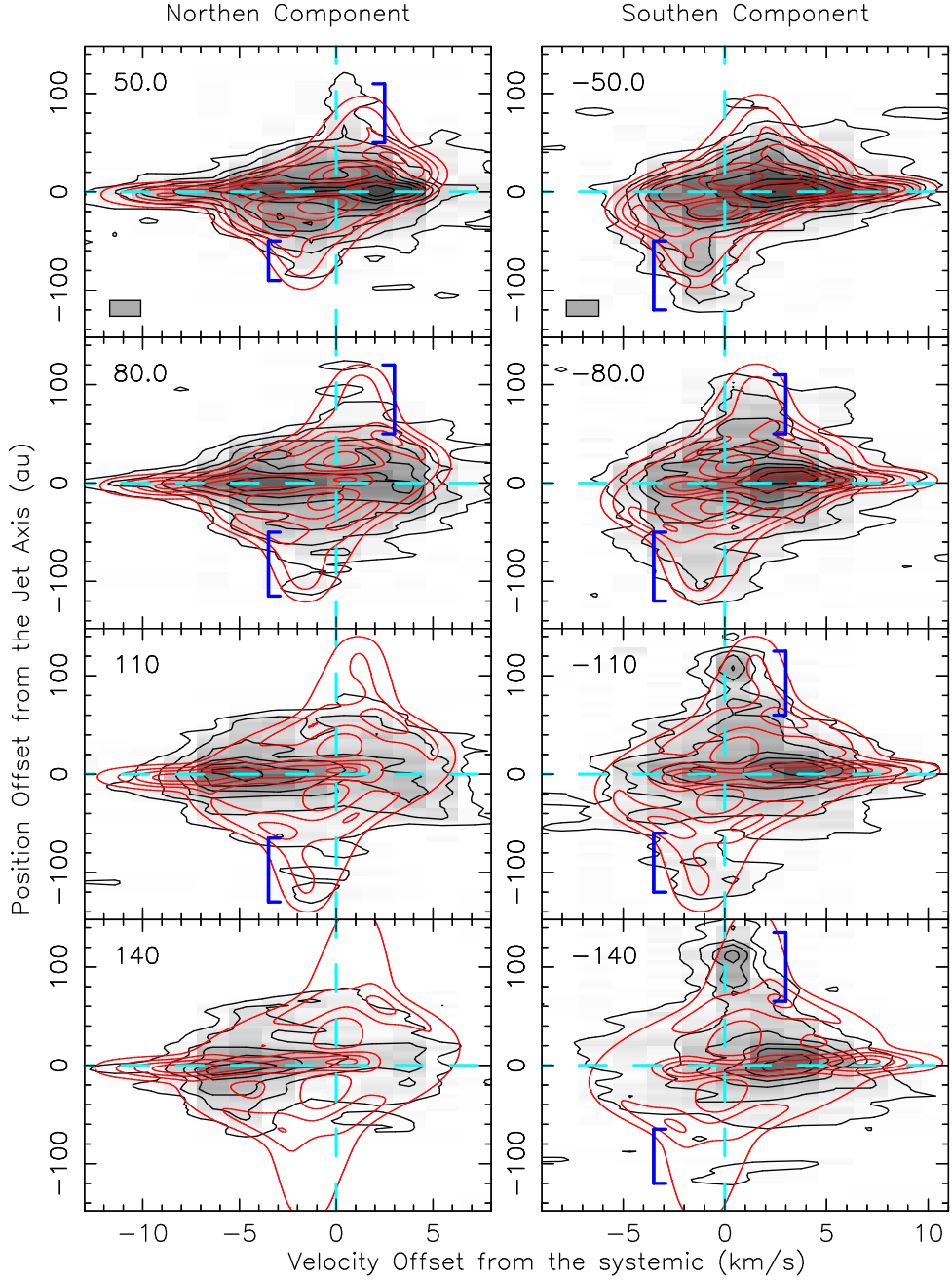
 \centering 
\putfiga{0.7}{0}{f4.eps} 
\figcaption[]
{Comparison of model PV diagrams (red contours) to the 
observed PV diagrams (black contours) of SO emission
cut across the jet axis centered at increasing distances (as indicated in the upper left corners in au)
from the protostar along the jet axis.
Contours start at 2$\sigma$ with a step of 3$\sigma$, where $\sigma \sim 4.5$ K.
Blue brackets mark the regions where the unperturbed wide-angle outflow is detected.
The bright emission near the jet axis is from the jet.
The emission in between the jet and the unperturbed wide-angle outflow
is from the shells.
\label{fig:pvSOBensjet}}
\end{figure}

Figure \ref{fig:pvSOBensjet} shows the resulting model PV diagrams on the
observed ones cut across the jet axis centered at increasing distance from
the protostar to near the end of the wide-angle outflow, with $\xSO \sim
4.5\times10^{-7}$ in the extended disk wind and $\xSO \sim 6.2\times10^{-5}$
in the jet.  As can be seen, with the outer unperturbed part of the wind,
this model roughly reproduces the PV structures of the faint unperturbed
wide-angle outflow (marked with blue brackets), even though it predicts a
rotation velocity slightly larger than observed.  Note that the wide-angle
outflow in the north is only detected within $\sim$ 130 au of the protostar
(see also Figure \ref{fig:SOmom1}). Moreover, with the inner part of the wind
being radially expanding, the model produces tilted elliptical PV structures
for the shell and can roughly match the observations with the dynamical age
$\ts \sim 37$ yrs, similar to that found in \citet{Lee2018Dwind}.  The
resulting outflow velocity is drawn in Figure \ref{fig:cartoon}.  This age
roughly agrees with the axial distance ($\sim 600$ au) traveled by the
cavity apex for a jet speed $\sim$ 64 \vkm{}, supporting that the shell at
the base is created by the same large jet bowshock seen downstream. 
However, since the model PV structures of the shell are tilted more than
observed, the rotation velocity in the shell is also over-predicted.  Note
that the density in the shell is $\sim$ 4 times that in the unperturbed
disk-wind model in order to match the observed SO intensity there.  The
model can produce linear PV structures for the jet with a broad range of
velocities near the jet axis, also roughly consistent with the observations.



\begin{figure}
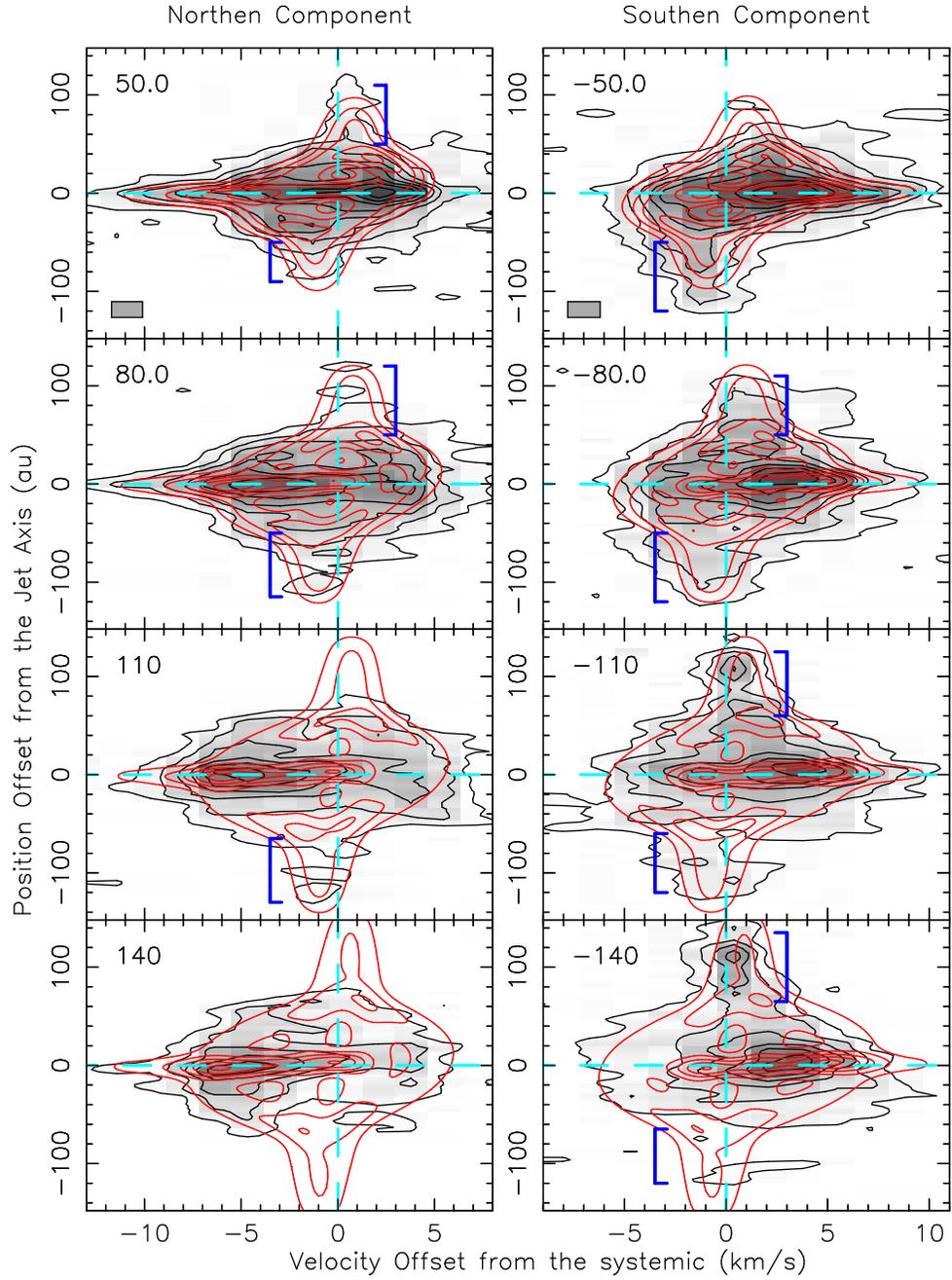
 \centering 
\putfiga{0.7}{0}{f5.eps} 
\figcaption[]
{Same as Figure \ref{fig:pvSOBensjet} but with a smaller $\lambda \sim 3.5$ in order to better reproduce the rotation
velocity of the wide-angle outflow including the shells.
\label{fig:pvSOsjet}}
\end{figure}

In order to obtain a better match to the observed rotation in the wide-angle
outflow, we reduce the rotation velocity and poloidal velocity in the model
by lowering the $\lambda$ value, keeping the same for the other parameters. 
At the same time, we decrease the inner launch radius of the SO jet to 0.05
au to maintain the same maximum jet velocity.  As shown in Figure
\ref{fig:pvSOsjet}, this very simple ``modified'' model with
$\lambda \sim 3.5$ provides a better match to the observations.  Such a low
$\lambda$ value is favored by recent MHD simulations including stellar
irradiation \citep{Wang2019}.  Moreover, with this smaller $\lambda$,
the wind density will be a factor of $\sim$ 3 higher and thus the required
SO abundance will become $\simeq 10^{-7}$, close to that in the HH 212 disk
\citep{Podio2015}. This result is in excellent agreement with
thermo-chemical modeling of dense Class 0 MHD disk winds
\citep{Panoglou2012}, which predicts that the SO wind abundance should
remain ``frozen" near the disk value up to $z \simeq 100$ au.  The sudden
drop of the wind SO brightness observed above $z \sim 150$ au (Figures
\ref{fig:SiO_SO} and \ref{fig:SOmom1}) is also predicted by this model, as
the wind becomes transparent to photodissociating FUV photons from the
accretion shock \citep{Panoglou2012}.  The disk wind is thus dense enough to
remove the disk angular momentum, if its SO abundance is close to that in
the disk.

\section{Conclusions}

Our new observations have reconciled previous studies of SO rotating outflow
in HH 212, supporting the presence of an extended MHD disk wind out to the
disk outer edge at 40 au, removing most of the angular momentum flux
required for accretion \citep{Tabone2017,Tabone2020}, but with a smaller
magnetic level arm and an inner launching radius truncated to $\gtrsim$ 4
au.  This extended wind is likely a steady wind continuously ejected
from the disk.  However, the SO and SiO jet is resolved as a chain of knots,
most likely caused by time variations in the jet ejection velocity
\citep{Eisloeffel1992}.  In particular, three different periods of ejections
have been estimated for the knots and bowshocks \citep{Lee2020Jet}.  This
could in turn be due to episodic accretion near the innermost disk,
triggered by gravitational instability \citep{Vorobyov2018}, a binary
companion, and/or a star-disk interaction.  The jet from the innermost disk
($\ro \sim 0.05-0.20$ au) drives large bow shocks interacting with the
extended disk wind and producing a cavity, with the thin SO shell forming
its boundary. These data are thus providing the first unambiguous evidence
for the theoretically predicted interaction between a time-variable jet and
an outer disk wind \citep{Tabone2018}.  Resolving such an interaction is
crucial for the proper interpretation and modeling of less well resolved
disk wind candidates.

Furthermore, the width of the cavity provides us with the first quantitative
 clues to the magnetic field strength in a disk wind.  Indeed,
the twisting of field lines at the base of MHD disk winds creates a strong magnetic 
pressure that efficiently confines the sideways expansion of jets and jet 
bow shocks \citep{Meliani06,Matsakos09}. 
With a jet mass-flux $\sim 10^{-6}$ \solarmass{} yr$^{-1}$
ejected sideways at a speed $\sim$ 10 \vkm{} \citep{Lee2015}, 
and the magnetic field strength required in our wind model to drive accretion at
$\dot{M}_{in}\sim 3\times10^{-6}$ \solarmass{}yr$^{-1}$,
equilibrium at $z \sim 150$ au between the lateral ram pressure
and magnetic pressure 
is reached on the streamline launched from 4 au, as observed. Hence, 
the cavity width appears consistent with the outer disk ($r_0 \ge 4$ au) 
hosting a large-scale magnetic field sufficient to extract angular momentum 
at the observed rate.  In addition, since magnetic pressure in self-similar MHD 
disk winds increases inward faster ($\propto r_0^{-5/2}$) than the 
wind ram pressure ($r^{-2}$), the disk wind 
should be much weaker inside of $r_0 \sim 4$ au, otherwise 
the bowshock could not have expanded that far.  MHD numerical simulations 
of the jet/disk-wind interaction in HH 212 will be able to confirm and refine these results.

In the innermost disk with a radius of $\sim$ 0.2 to 4 au, another mechanism
such as magneto-rotational instability (MRI) \citep{Balbus2006} might thus
be needed to transport disk angular momentum outward, allowing material to
accrete to the innermost radius to the dust-free zone .  Interestingly, 4 au
is roughly of the same order as the predicted transition radius between the
ionized inner disk (where MRI can grow) and the low-ionization external
``non-ideal zone" (where the MRI is quenched by non-ideal MHD effects)
\citep{Armitage2011,Mori2019}.  Inside this radius, the disk surface
temperature in HH 212 becomes $\gtrsim 1000$ K \citep{Lee2017Disk} and thus
MRI could turn on.  A transition from an MHD disk-wind regime outside to an
MRI-dominated regime inside is also predicted when disk magnetization
decreases inward, due to, e.g., magnetic flux diffusion
\citep{Ferreira2013}.  If confirmed, the transition suggested by our
observations would then bring unique clues to ionization and field diffusion
in young disks.




\acknowledgements

This paper makes use of the following ALMA data:
ADS/JAO.ALMA\#2017.1.00044.S.  ALMA is a partnership of ESO (representing
its member states), NSF (USA) and NINS (Japan), together with NRC (Canada),
NSC and ASIAA (Taiwan), and KASI (Republic of Korea), in cooperation with
the Republic of Chile.  The Joint ALMA Observatory is operated by ESO,
AUI/NRAO and NAOJ.  C.-F.L.  acknowledges grants from the Ministry of
Science and Technology of Taiwan (MoST 107-2119-M-001-040-MY3) and the
Academia Sinica (Investigator Award AS-IA-108-M01).  B.T.  acknowledges
funding from the research programme Dutch Astrochemistry Network II with
project number 614.001.751, which is financed by the Dutch Research Council
(NWO).

\clearpage

\end{document}